\documentclass[twocolumn]{emulateapj}

\def\alwaysmath#1{\ifmmode{#1}\else{$#1$}\fi}
 
\lefthead{Law et al. 2005}
\righthead{Strategies for Integral-Field Spectroscopy}

\begin{document}
\title{Predictions and Strategies for Integral-Field Spectroscopy of High-Redshift Galaxies}
\author{\sc David R. Law, Charles C. Steidel, and Dawn K. Erb}
\affil{California Institute of Technology, Department of Astronomy, MS 105-24,
Pasadena, CA 91125 (drlaw, ccs, dke@astro.caltech.edu)}

\begin{abstract}
We investigate the ability of infrared integral-field spectrographs to map the velocity fields of high redshift galaxies,
presenting a formalism which may be applied to any telescope and imaging spectrograph system.  
We discuss the 5$\sigma$ limiting line fluxes which current integral-field spectrographs will reach, and extend this
discussion to consider future large aperture telescopes with cryogenically cooled adaptive reimaging optics.
In particular, we simulate observations of spectral line emission from
star-forming regions at redshifts $z = 0.5$ to 2.5
using a variety of spatial sampling scales and give predictions for the signal-to-noise ratio expected
as a function of redshift.
Using values characteristic of the W.M. Keck II telescope and the new OH-Suppressing InfraRed Imaging Spectrograph (OSIRIS) we calculate
integral-field signal-to-noise ratio maps for a sample of $U_nG{\cal R}$ color-selected star-forming galaxies at redshift $z \sim 2 - 2.6$ and
demonstrate that OSIRIS will be able to reconstruct the two-dimensional projected 
velocity fields of these galaxies on scales of 100 mas ($\sim$ 1 kpc at redshift $z \sim 2$).
With signal-to-noise ratios per spatial sample up to $\sim$ 20,
OSIRIS will in some cases be able to distinguish between merger activity and ordered disk rotation.  Structures on scales smaller than 1 kpc 
may be detected by OSIRIS for particularly bright sources, and will be easy targets for future 30m class telescopes.

\end{abstract}


\section{INTRODUCTION}
As the available instrumentation on large telescopes has pushed observations into the near-infrared (NIR), the denizens of the so-called ``redshift desert'' from
$z \sim 1.4 - 2.5$ have increasingly become targets of concerted study.
Observations suggest that the morphological chaos of the $z \sim 3$ universe evolves into the familiar modern-day Hubble sequence by
redshift $z \sim 1$ (e.g. Giavalisco et al. 1996, Papovich et al. 2005), and the redshift $z \sim 2$ population is likely to contain a wide assortment of
galactic types as protogalactic merging activitity transitions to kinematically ordered evolution.
It often proves difficult however to distinguish starbursting galaxies with an ordered disk morphology
from those actively experiencing protogalactic assembly since the relevant angular scales are barely resolved under typical ground-based seeing conditions, and the rest-frame
UV light (e.g. as observed in the optical by the Hubble Space Telescope) is dominated by
emission from bright, clumpy star-forming regions.

Long-slit nebular emission line spectroscopy has enabled velocity curves to be derived for a sample of high-redshift galaxies (e.g. Pettini et al. 2001, Lemoine-Busserolle
et al. 2003, Moorwood et al. 2003, Erb et al. 2003, Erb et al. 2004), giving some insight into the kinematic properties of these sources and suggesting
that massive disk systems may be present out to at least redshift 2.5.  However, ground-based long-slit spectroscopy suffers from severe
atmospheric seeing
problems (as demonstrated by Erb et al. 2004), and recovered velocity curves for morphologically selected potentially
disk-like galaxies at redshift $z \sim 2$ often show little to no spatially resolved velocity shear (Erb et al. 2004).
One possible explanation for this lack of velocity shear may be that these sources are composed of
proto-galactic fragments of angular size below the typical seeing limit
($\sim 0\farcs5$ in the $K$-band).

Given the small
apparent angular size of these sources (radius $r \lesssim 1\farcs0$) compared to that of the seeing disk ($\Theta_{\rm seeing} \approx 0\farcs5$), 
high spatial resolution data from ground-based
telescopes has only recently become available through the aid of adaptive-optics (AO) technology.
The advent of integral-field spectrographs (IFS) on 10m-class telescopes equipped with laser-guide star adaptive optics systems offers the new possibility of mapping the
kinematics of galaxies at $z \sim$ 2 on the scale of tenths of an arcsecond or less.  Not only can such AO-fed spectrographs mitigate
the seeing-induced uncertainties inherent in velocity curves derived from long-slit spectra, but their integral-field capability will provide two-dimensional
velocity maps from which it may be possible to reconstruct the kinematic profiles of these galaxies on scales of a few hundred parsecs.

Recent years have seen a rapid increase in the number of optical/NIR integral-field spectrographs capable of studying the velocity fields of high redshift galaxies
(Table \ref{speclist.table}), and these instruments are starting to yield kinematic maps of large galaxies from redshifts $z =$ 0.1 (Swinbank et al. 2005a) and 0.5
(Flores et al. 2004) up to $z =$ 2.4 (Swinbank et al. 2005b).
In this contribution, we outline an observing strategy for integral-field IR spectroscopy of high redshift galaxies, treating the case
of a general telescope + IFS system in \S 2.
In \S 3 we narrow our discussion to concentrate on simulating the capabilities of the OSIRIS spectrograph on the W.M. Keck II telescope,
and predict the signal-to-noise (S/N) ratios expected for observations of nebular line emission from star-forming regions with redshifts
from  $z \sim 0.5 - 2.5$.
We explore the relative merits of available lenslet scales and observable spectral emission lines, 
and assess the ability of OSIRIS to obtain spatially resolved velocity information from $U_nG{\cal R}$ color-selected
galaxies at redshift $z \sim 2$ - 2.5 in the Hubble Deep Field North (i.e. the ``BX'' sample
defined by Steidel et al. 2004 and Adelberger et al. 2004).
Finally, in \S 4 we use characteristics typical of the OSIRIS spectrograph to investigate the limiting line fluxes detectable from an arbitrary source
using current generation spectrographs, extending this discussion to consider next-generation telescopes and cryogenically cooled optical systems.

\begin{deluxetable*}{lccccc}
\tablecolumns{6}
\tablewidth{0pc}
\tabletypesize{\footnotesize}
\tablecaption{Current Optical/IR Integral-Field Spectrographs}
\tablehead{
\colhead{Spectrograph} & \colhead{Telescope} & \colhead{Optical\tablenotemark{a}} & \colhead{NIR\tablenotemark{b}} & \colhead{AO\tablenotemark{c}} & \colhead{Reference}}
\startdata
GIRAFFE & VLT & yes & no &  no& Hammer et al. (1999)\nl
GMOS & Gemini & yes & no & NGS & Davies et al. (1997)\nl
GNIRS & Gemini & no & yes & no & Allington-Smith et al. (2004)\nl
INTEGRAL & WHT & yes & no & NGS & Garc{\'{\i}}a-Lorenzo et al. (2000)\nl
NIFS & Gemini & no & yes & NGS & McGregor et al. (2003)\nl
OSIRIS & Keck II & no & yes & LGS & Larkin et al. (2003)\nl
SAURON & WHT & yes & no & no & Bacon et al. (2001)\nl
SINFONI/SPIFFI & VLT & no & yes & LGS & Tecza et al. (1998)\nl
UIST & UKIRT & no & yes & no & Ramsay Howat et al. (1998)\nl
\enddata
\tablenotetext{a}{Wavelength coverage at least $\lambda\lambda$ 4000 - 7000 \AA.}
\tablenotetext{b}{Wavelength coverage at least $\lambda\lambda$ 1 - 2.5 \micron.}
\tablenotetext{c}{Present adaptive-optics capability; natural guide star (NGS) or laser guide star (LGS).}
\label{speclist.table}
\end{deluxetable*}

We assume a cosmology in which $h = 0.7$, $\Omega_m = 0.3$, $\Omega_{\Lambda} = 0.7$.

\section{METHOD}
Integral-field (or multi-object) spectrographs may have optical designs ranging from a simple objective prism to a more
complex fiber-fed, lenslet-based, or image-slicing configuration. We tailor our discussion specifically to the case of an imaging lenslet array,
in which each of a closely spaced array of lenslets disperses a spectrum in a staggered pattern onto a detector (a good introduction to this
technique is given by Bacon et al. 1995).  Our formalism is presented with such a lenslet-type spectrograph in mind, but
may be easily modified for both image-slicing and fiber-type spectrographs.

\subsection{Signal Estimation}
\subsubsection{Estimating Background Count Rates}
We adopt the Gemini Observatory model for the Mauna Kea near-IR sky brightness spectrum\footnote{Available on the Web at 
http://www.gemini.edu/sciops/ObsProcess/obsConstraints/atm- models/nearIR\_skybg\_16\_15.dat} (spectral resolution ${\cal R} \sim 2000$)
which incorporates zodiacal emission (from
a 5800 K blackbody), atmospheric emission (from a 250 K blackbody), and radiation from atmospheric OH molecules.
In addition to these components
we introduce models of the thermal blackbody emission from warm reflective surfaces in the light path.
The telescope mirrors and the adaptive reimaging optics are modelled as 270 K blackbodies with net emissivities
$\epsilon = 1 - \eta$ (where $\eta$ is the transmission coefficient of the component).

Using the Planck blackbody function, the photon flux due to thermal emission from these components is
\begin{equation}
f_{\lambda} = \epsilon\, \alpha^2\, \frac{2\, c}{\lambda^4}\frac{1}{e^{hc/\lambda k T}\,-\,1} \,{\rm photons}\,{\rm s}^{-1} \,{\rm cm}^{-2} \,{\rm cm}^{-1} \,{\rm arcsec}^{-2}
\label{bbeqn}
\end{equation}
where $f_{\lambda}$ is the spectral flux at wavelength $\lambda$,
$\alpha = 4.85 \times 10^{-6}$ is a scale factor representing the number of steradians per square arcsecond,
$T$ is the temperature of the component in degrees Kelvin, and the remaining symbols $c,h$ and $k$ are standard physical constants.

The corresponding total electron count rate per detector pixel per lenslet
($R_{\rm BG}$, electrons s$^{-1}$ lenslet$^{-1}$ pixel$^{-1}$) from these three background sources (Gemini sky brightness model, telescope
emissivity, and AO system emissivity)
is given by
\begin{equation}
R_{\rm BG}(\lambda) = (f_{\lambda, {\rm sky}}\, \eta_{\rm tel}\, \eta_{\rm AO} + f_{\lambda, {\rm tel}} \eta_{\rm AO} + f_{\lambda, {\rm AO}})\, \eta_{\rm spec}(\lambda) A\, a^2 \frac{P}{{\cal N}_1}
\label{bgrateequation}
\end{equation}
where $f_{\lambda, {\rm sky/tel/AO}}$ represents the photon flux from the sky/telescope/AO system respectively in photons s$^{-1}$ cm$^{-2}$ cm$^{-1}$ as$^{-2}$, 
$\eta_{\rm tel}$ is the throughput of the telescope optics, $\eta_{\rm AO}$ is the throughput of the AO system, $\eta_{\rm spec}(\lambda)$ is the throughput of the
spectrograph (a complicated function of $\lambda$ including the throughput of the spectrograph blaze function, filter transmission, detector response, and optical throughput
additional components),
$A$ is the telescope collecting area, $a$ is the angular lenslet size (in arcseconds), $P$ is the pixel scale (wavelength coverage
per pixel), and ${\cal N}_1$ is the spatial width of each spectrum in detector pixels.

\subsubsection{Estimating Source Count Rates}
The source count rate in a given emission line may be estimated by scaling a finely sampled (i.e. with spatial resolution at least as fine as the desired lenslet
scale) map of the source morphology by
the total emission line flux of the source.
While H$\alpha$ maps
of high redshift galaxies are generally unavailable,
H$\alpha$ emission often correlates well with the FUV (1500\AA) morphology,
as demonstrated for late-type local galaxies by Gordon et al. (2004), Kennicutt et al. (2004), and Lee et al. (2004),
and it is often a reasonable approximation to use a FUV flux map to represent the H$\alpha$ flux distribution (although cf. Conselice et al. 2000).
The FUV flux can readily be observed by the Hubble Space Telescope's Advanced Camera for Surveys (HST-ACS) since rest-frame FUV emission is redshifted
into the $B$-band for sources at redshift $z \sim 2 - 2.6$, and we therefore use images of our target galaxies in the GOODS-N field which are publically available
from the HST-ACS data archives\footnote{Based on observations 
made with the NASA/ESA Hubble Space Telescope, obtained from the data archive at the Space Telescope Science 
Institute. STScI is operated by the Association 
of Universities for Research in Astronomy, Inc. under NASA contract NAS 5-26555.}.  These images are ideal for 
studying the morphology of the target galaxies on angular scales of less than an arcsecond since the point-spread
function of the ACS data is about 125 mas (Giavalisco et al. 2004).

These FUV flux maps are resampled into pixel maps with a pixel size corresponding to the desired lenslet scale, and the sum of the
pixel values is rescaled by the total line flux of the source divided by the average photon energy
to obtain a grid of line fluxes $f_{\lambda, {\rm source}}$ (photons s$^{-1}$ cm$^{-2}$ lenslet$^{-1}$)
for each lenslet.
Since the adopted ACS flux maps have already been sky-subtracted we assume that all pixels within a given region with fluxes above zero-threshold
represent source emission, and find the total FUV source flux by simply summing all such pixels.
The average electron count rate per detector pixel (electrons s$^{-1}$ lenslet$^{-1}$ pixel$^{-1}$)
is then calculated in analogy to Equation \ref{bgrateequation} as
\begin{equation}
R = f_{\lambda,{\rm source}}\, \eta_{\rm atm}(\lambda) \, \eta_{\rm tel}\, \eta_{\rm AO}\, \eta_{\rm spec}(\lambda) \, \frac{A}{\Delta \lambda}\, \frac{P}{{\cal N}_1}
\label{rateequation}
\end{equation}
where $\Delta \lambda$ is the full width at half maximum (FWHM) of the spectral line (computed as the quadrature sum of the line width due to source
velocity dispersion $\sigma_{\rm v}$ and from
the finite spectrograph resolution) and $\eta_{\rm atm}(\lambda)$ is the atmospheric transmission as a function of wavelength, for which
we adopt the Mauna Kea transmission spectra generated by Lord 1992.\footnote{Both the Mauna Kea transmission spectra and sky brightness profiles assume
1.6 mm H$_2$O and sec($z$) = 1.5 at local midnight.}

\subsubsection{Adaptive-Optics and Seeing Corrections to Count Rates}
We now take into account the effects of realistic AO performance and losses to the atmospheric seeing halo (the net throughput of the AO system was already
taken into account in \S 2.1.1 and 2.1.2).
The flux in each lenslet is resampled uniformly into a grid of $x$ $\times$ $x$ 
squares\footnote{$x$ is chosen to be suitably large that the results of this method converge, but
small enough that computational efficiency is reasonably high.  We find that $x = 5$ gives a good convergence of the PSF integration with a reasonable efficiency
for all except the finest sampling scales
of $a= 20$ mas, for which we employ $x=2$.  Note that the efficiency will depend also on the total size of the field covered by the lenslet array.}, 
and the approximation made that
the flux in each grid square (before incidence upon the Earth's atmosphere) may be described as 
a point source.  After traversing both the atmosphere and the AO system,
the light from this point source is divided between a diffraction-limited core and a seeing-limited halo, the fluxes of which
are given by
\begin{eqnarray}
f_{\rm diffraction} =& {\cal S} \times f_{\rm total}\\
f_{\rm seeing} =& (1 - {\cal S}) \times f_{\rm total}
\label{strehleqn}
\end{eqnarray}
where ${\cal S}$ is the strehl of the AO correction.  Assuming that the net wavefront error of the adaptive optics
system $\sigma_{\rm wfe}$ is constant, strehl\footnote{The strehl
actually achieved in laser-guide star operation for a given source will also depend 
on the proximity of the source to a suitably bright (magnitude $R < 18$) reference star for tip-tilt
correction of the wavefront.}
as a function of wavelength is given by
\begin{equation}
{\cal S} = e^{-(2 \pi \sigma_{\rm wfe} / \lambda)^2}
\end{equation}

Both the seeing-limited halo and diffraction-limited core may be approximately represented by two-dimensional Gaussian profiles
with $\sigma_{\rm halo} = \frac{\theta_{\rm seeing}}{2.4}$ and
$\sigma_{\rm core} = \frac{1.22 \,\lambda/D}{2.95}$ (where $D$ is the diameter of the telescope's primary mirror).\footnote{The point spread function (PSF) of the
diffraction-limited core is best
described by a two-dimensional Airy function, which has a central peak narrower than that of a standard Gaussian with the same peak value.  
However, the 2D Airy function profile
is closely approximated by 
a 2D Gaussian whose FWHM has been reduced by a factor of $2.4/2.95 = 0.81$.  Since the exponential form of the Gaussian
lends itself more readily to analytic integration than the Airy function,
in the interests of efficiency
we adopt this Gaussian approximation.}
The core and halo light profiles from each grid square may be integrated over each lenslet to generate an observed flux
distribution which incorporates both {\it losses to} the seeing halo from a given lenslet and {\it gains from} the seeing halos
of neighboring lenslets.

Since this procedure assumes that light lost to the seeing halo of one lenslet is recovered by surrounding lenslets, 
it is possible that all losses to the seeing halo from a given
lenslet may be regained from spill-over flux from the halos of neighboring lenslets if the intrinsic flux in these neighbors is suitably high.
However, this flux will generally be a source of additional background noise - if the flux is spilled from a region of the source
with a very different line-of-sight velocity it will contribute this flux at the corresponding wavelength and hinder precise measurement
of the spectral peak of flux originating from a given lenslet.  Modeling this complex effect in detail requires {\it a priori} knowledge
of the velocity substructure of the source, and we therefore generally assume that all such spilled flux adds to the noise of the observation rather than the signal.
In \S 2.2 and \S 3.4 where we adopt specific velocity models however, spilled light is treated comprehensively as a function of both angular
position and wavelength.

\subsubsection{Estimating Signal-to-Noise Ratios}
Using the quantities derived it is possible to calculate the S/N ratio expected for detection
of a particular spectral line in each lenslet.  We use the standard equation
\begin{equation}
\frac{S}{N} = \frac{R \,{\cal N}_1 \,{\cal N}_{\rm spec} \,t}{\sqrt{(R+R_{\rm BG}+R_{\rm DK})\,{\cal N}_1 \,{\cal N}_{\rm spec} \,t + N_{\rm RD}^2\,(t/t_0)\,{\cal N}_1 \,{\cal N}_{\rm spec}}}
\label{SNequation}
\end{equation}
where ${\cal N}_{\rm spec}$ is the number of spectral pixels summed (i.e. the FWHM of the spectral line in pixels), $R_{\rm DK}$ is the dark
current count rate in e$^-$ s$^{-1}$ pixel$^{-1}$, $N_{\rm RD}$ is the detector read noise in e$^-$ pixel$^{-1}$,
$t$ and $t_0$ are the total and individual exposure times respectively, and all other quantities have been previously defined.
%

\subsection{Spectral Synthesis}
Simply {\it detecting} an emission line with some confidence does not guarantee that its {\it centroid} can be measured
with great accuracy, and we therefore generate
artificial spectra and attempt to measure their centroids using techniques which will be applied to observational data
to effectively simulate the recovery of velocity substructure.

Defining a spectral vector with one element for each spectral pixel (i.e. summing over the ${\cal N}_1$ detector pixels to create a 1-dimensional spectrum),
it is easy to generate a simulated spectrum from the counts as a function of wavelength by introducing noise according to Poisson statistics.
Using this simulated spectrum (or an observed spectrum which has been precisely wavelength calibrated using the bright OH spectral features),
the background signal $N_{BG} = (R+R_{\rm BG}+R_{\rm DK})\,{\cal N}_1 \,{\cal N}_{\rm spec} \,t + N_{\rm RD}^2\,(t/t_0)\,{\cal N}_1 \,{\cal N}_{\rm spec}$
due to read noise, 
dark current, and the sky background may be subtracted off\footnote{Generally this background signal will be well-determined 
for all except the smallest fields of view.  We explore this for the case
of the OSIRIS spectrograph in \S 3.1.} to determine the sky-subtracted counts as a function 
of wavelength for the source.  This spectrum may be further processed to obtain velocity centroids
using standard routines.  Hand-processing by an experienced individual appears to be the most reliable method
of measuring velocity centroids, although such a method is impractically time-consuming for extensive simulations with hundreds of spectra apiece.
Instead, we adopt
an automated routine which fits a gaussian profile to the background-subtracted spectra using a Levenberg-Marquardt fitting algorithm
(a non-linear least-squares regression algorithm similar to that used in the standard IRAF routine {\it splot}) and measures the velocity shift of the
peak flux from the systemic redshift.  This method is found to perform reliably for most spectral features with
S/N ratios $> 3$, although some particularly faint features which are not handled well by this automated method may still yield useful information
when individually hand-processed.

We note that for the case of spectrographs whose output is in the form of data cubes, one-dimensional observational spectra may be extracted for each spatial element,
and may then be processed in entirely the same manner as these simulated spectra.

%

\section{SIMULATED OBSERVATIONS WITH OSIRIS}

The simulation method described in the previous section has been outlined in a general manner 
so that it may be applied to any telescope and integral-field spectrograph system.  
We now proceed to adopt characteristics typical of the W.M. Keck II telescope and OSIRIS spectrograph in order to quantify the performance capability
expected from this system for the study of high-redshift galaxies.  In \S 3.1 we describe the site, telescope, 
and spectrograph characteristics, and in \S 3.2 we explore the
expected detection limits of
OSIRIS for observations of nebular line emission from typical star-forming regions at redshifts $z = 0.5 - 2.5$.
In \S 3.3 we present integral-field S/N ratio maps for selected $z \sim 2 - 2.6$ star-forming galaxies
in the GOODS-N field and explore the integration times and lenslet scales required to
achieve significant detections.  Finally, in \S 3.4 we explore the ability of OSIRIS to recover the velocity substructure of these sources
assuming disparate kinematic models.  We note that the assumed characteristics of the OSIRIS spectrograph are preliminary results
based upon commissioning data, and the results of our analysis should therefore be scaled accordingly to reflect the
final well-determined specifics of the instrument once such information is available.

\subsection{The OSIRIS Spectrograph}

\subsubsection{Mechanical Overview}

OSIRIS (OH-Suppressing InfraRed Imaging Spectrograph) is an integral-field spectrograph designed and built at UCLA for the Keck II telescope (Larkin et al. 2003), 
and is presently in the initial stages of commissioning
during the 2005 observing season.  The spectrograph employs reimaging optics to select between lenslet scales of 20, 35, 50, and 100 mas and uses a low dark-current
Hawaii-2 HgCdTe 2048 $\times$ 2048 IR detector array.  OSIRIS is designed to work in conjunction with the Keck-II laser guide star (LGS) AO system since
the vast majority of targets (about $\sim 99$\% of the sky) are too distant from a suitably bright star to permit natural guide star (NGS) observations.
Characteristic parameters for the Keck II telescope, LGS AO system, and OSIRIS spectrograph are summarized in Table \ref{Detector.table}.
The spectrograph parameters given are based upon values provided in the OSIRIS pre-ship report (J. Larkin, private communication).

\begin{deluxetable}{lcc}
\tablecolumns{3}
\tablewidth{0pc}
\tabletypesize{\scriptsize}
\tablecaption{Keck/OSIRIS System Characteristics}
\tablehead{
\colhead{Parameter} & \colhead{Symbol} & \colhead{Value}}
\startdata
\multicolumn{3}{c}{Telescope/AO characteristics}\nl
\hline
AO throughput\tablenotemark{a} & $\eta_{\rm AO}$ & 65\%\nl
Collecting area & $A$ & $7.85 \times 10^5$ cm$^2$\nl
$K$-band diffraction limit & $\theta_{\rm diff}$ & 55 mas\nl
$K$-band seeing & $\theta_{\rm seeing}$ & 550 mas\nl
Telescope throughput\tablenotemark{a,b} & $\eta_{\rm tel}$ & 80\%\nl
Wavefront error\tablenotemark{c} & $\sigma_{\rm wfe}$ & 337 nm\nl
\cutinhead{OSIRIS Characteristics\tablenotemark{a}}
Average OSIRIS throughput & $\eta_{\rm spec}$ & 15\% ($z$-$H$ bands)\nl
 & & 20\% ($K$ band)\nl
Average system throughput\tablenotemark{d} & $\eta$ & 8\% ($z$-$H$ bands)\nl
 & & 10\% ($K$ band)\nl
Dark current & $R_{\rm DK}$ & .03 e$^-$ s$^{-1}$ pixel$^{-1}$\nl
Grating blaze & \nodata & 6.5 $\micron$\nl
Read noise\tablenotemark{e} & $N_{\rm RD}$ & 3 e$^-$ pixel$^{-1}$\nl
Spectral length & ${\cal N}_2$ & 1600 pixels (broadband)\nl
 & & 400 pixels (narrowband)\nl
Spectral resolution & ${\cal R}$ & 3900 (20, 35, 50 mas lenslets)\nl
 & & 3400 (100 mas lenslets)\nl
Spectral width & ${\cal N}_1$ & 3 pixels\nl
\enddata
\tablenotetext{a}{Values based on OSIRIS pre-ship report and preliminary commissioning results (J. Larkin, private communication).}
\tablenotetext{b}{Assumes 90\% throughput of primary and secondary mirrors.}
\tablenotetext{c}{Based on achieved $K$-band strehl of 0.36 (http://www2.keck.hawaii.edu/optics/lgsao/performance.html)}
\tablenotetext{d}{Average system throughput $\eta = \eta_{\rm tel}\eta_{\rm AO}\eta_{\rm spec}$}
\tablenotetext{e}{After multiple-correlated double samples.}
\label{Detector.table}
\end{deluxetable}

OSIRIS uses broad- and narrow-band order-sorting filters to select the wavelength range imaged on the detector; the available four broadband IR filters 
(outlined in Table \ref{Filters.table}) are chosen
to correspond to the regions of high throughput in orders $n = 3 - 6$ of the spectrograph grating (blazed at 
6.5 \micron).
Operating in broad-band mode, OSIRIS is capable of producing spectra from 16 x 64 lenslets
for a total of 1024 spectra with 1600 spectral pixels each.  OSIRIS is also capable of operating in narrow-band mode using 19 narrow-band filters which together
span the same range of wavelengths as the broadband filter set.  This narrow-band mode will generally be more useful for targeted velocity studies of
high-redshift galaxies, for which velocity-shifted nebular emission lines are expected to fall within a few angstroms of the systemic value.
In this mode, spectra from 48 x 64 lenslets can fit on the detector for a total of 3072 spectra comprised of 400 spectral pixels each.  
The resulting fields of view for broadband and narrow-band operation at each
of the four available lenslet scales are given in Table \ref{FOV.table}.

\begin{deluxetable}{cccccc}
\tablecolumns{6}
\tablewidth{0pc}
\tabletypesize{\scriptsize}
\tablecaption{OSIRIS Broadband Filters and Effective Model Sky Surface Brightness}
\tablehead{
\colhead{Band} & \colhead{Wavelength ($\micron$)} & \colhead{$\eta_{\rm filter}$\tablenotemark{a}} & \colhead{$\mu_{\rm AB}$\tablenotemark{b}} 
& \colhead{$\mu_{\rm Vega}$\tablenotemark{b}} & \colhead{$f_{\lambda}$\tablenotemark{c}}}
\startdata
$z$ & .98 --- 1.20 & 0.73 & 17.9 & 17.3 & 3.64044\nl
$J$ & 1.18 --- 1.44 & 0.81 & 17.8 & 16.8 & 3.33187\nl
$H$ & 1.47 --- 1.80 & 0.79 & 18.0 & 16.6 & 2.09761\nl
$K$ & 1.96 --- 2.40 & 0.81 & 15.2 & 13.3 & 20.5934\nl
\enddata
\tablenotetext{a}{Average narrowband filter throughput for filters within the indicated broadband wavelength range (J. Larkin, private communication).}
\tablenotetext{b}{Effective sky brightness in units of AB and Vega magnitudes per arcsecond$^{2}$.}
\tablenotetext{c}{Average effective background flux density in units of photons s$^{-1}$ nm$^{-1}$ arcsecond$^{-2}$ m$^{-2}$.}
\label{Filters.table}
\end{deluxetable}

In the interests of integration time, a suitably large field of view must be chosen so that a reasonable number of lenslets are imaging both the science target and 
the off-source background sky (for calibration purposes)
simultaneously.  In narrow-band mode, this means that at least 5\% of the lenslets must have negligible flux from the science target in order to 
calculate a background signal with an uncertainty of less
than 10\% of the Poisson shot noise.  Most redshift $z \sim 2$ star-forming galaxies have radial size $\sim$ 1 arcsecond
or less, resulting in a source area of
\begin{equation}
A_{\rm source} \lesssim \pi \left(\frac{1\, {\rm arcsecond}}{a}\right)^2\,\,\,{\rm lenslets}
\label{solidangleeqn}
\end{equation}
where $a$ is again the lenslet scale in arcseconds.  Operating at all except the finest sampling scales of $a = 20$ mas, these sources 
cover considerably less than the total number of lenslets
available over the field of view, indicating that accurate estimation of the background flux will be possible with no additional pointings.  
Studies using lenslet scales of $a = 20$ mas will require greater caution in
the placement of their fields of view to ensure reasonable background coverage, although as will be demonstrated in \S 3.3 these small sampling scales
are unlikely to be of great utility for OSIRIS observations of $z \sim 2$ galaxies anyway.

\begin{deluxetable}{ccc}
\tablecolumns{3}
\tablewidth{0pc}
\tabletypesize{\scriptsize}
\tablecaption{Available OSIRIS Angular Lenslet Scales and Fields of View\tablenotemark{a}.}
\tablehead{
\colhead{Lenslet Scale} & \colhead{Broadband FOV} & \colhead{Narrowband FOV}}
\startdata
0.02 & 0.32 $\times$ 1.28 & 0.96 $\times$ 1.28\nl
0.035 & 0.56 $\times$ 2.24 & 1.68 $\times$ 2.24\nl
0.05 & 0.8 $\times$ 3.2 & 2.4 $\times$ 3.2\nl
0.10 & 1.6 $\times$ 6.4 & 4.8 $\times$ 6.4\nl
\enddata
\tablenotetext{a}{All data are given in arcseconds and are based on the OSIRIS pre-ship report (J. Larkin, private communication).}
\label{FOV.table}
\end{deluxetable}

\subsubsection{Mauna Kea/OSIRIS Noise Characteristics}
\begin{figure}
\plotone{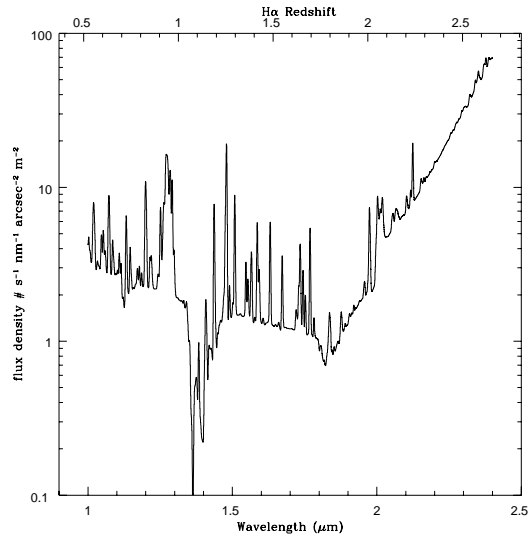}
\caption{Model effective background sky spectrum for Keck II incorporating the Gemini model for zodiacal light and atmospheric emission
(${\cal R}\, \sim$ 2000), 
a 35\% emissivity in the AO system, and a 20\% emissivity in the telescope optics (90\% throughput for both the primary and secondary mirrors)
at temperature $T = 270$ K.  Spectrally dense OH emission line regions appear as spikes since a spectrograph
with resolution ${\cal R} = 2000$ will not be able to reach inter-line background levels in these regions of severe contamination.  
The dip around 1.4 $\micron$ is due to high atmospheric opacity.  Redshifts
for which H$\alpha$ emission falls at a particular wavelength are indicated along the top axis of the plot.}
\label{background.fig}
\end{figure}

The model background sky brightness as a function of wavelength is calculated as outlined in \S 2.1.1 for typical Keck II and AO system characteristics
and is plotted in Figure \ref{background.fig}.  Note that for sake of clarity we suppress individual OH emission lines which are visibly separated from 
neighboring lines by a region of inter-line background, since it will in general be possible to work around these emission features.
Effective sky surface brightnesses corresponding to this background spectrum are given in Table \ref{Filters.table}.

It is informative to compute the relative contribution of detector read noise, dark current, and background radiation in determining the total noise in an exposure.
As given in Table \ref{Detector.table}, the read noise for the Hawaii-2 infrared detector array is $N_{\rm RD} \sim$ 3 electrons pixel$^{-1}$ 
(after multiple correlated double samples),
with a dark current of $R_{\rm DK} =$ 0.03 electrons s$^{-1}$ pixel$^{-1}$.  Using the model background spectrum (Fig. \ref{background.fig}), 
it is possible to calculate the minimum exposure time $t_{\rm min}$ required
for read noise to become a negligible contributor to the total noise budget,
\begin{equation}
t_{\rm min} = \frac{N_{\rm RD}}{R_{\rm DK} + R_{\rm BG}}
\end{equation}

\begin{figure}
\plotone{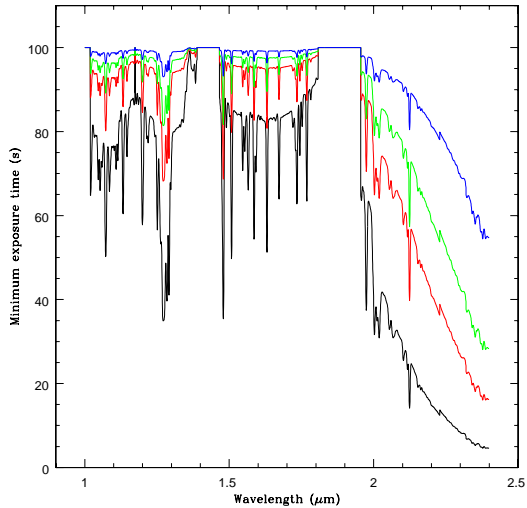}
\caption{Minimum integration time required for detector dark current $+$ background counts to equal the detector read noise for lenslet scales
of 100 (black line), 50 (red line), 35 (green line), and 20 mas (blue line).}
\label{mintime.fig}
\end{figure}

Figure \ref{mintime.fig} plots the minimum exposure time as a function of
wavelength for lenslet scales of 20, 35, 50, and 100 mas.  Note that for wavelengths 
where background radiation is negligible compared to the dark current $t_{\rm min}$ approaches 100 seconds
(since $N_{\rm RD}/R_{\rm DK} = $ 100 seconds),
while at wavelengths for which $t_{\rm min} < 50$ seconds the background radiation dominates over the dark current.
In order to minimize the contribution of detector read noise to 
our simulations we fix invidual exposure times at $t_0 =$ 900 seconds, although we note that shorter times could be chosen
on the basis of Figure \ref{mintime.fig} if long observations prove problematic due to technical difficulties
such as LGS-shuttering from beam collisions with other Mauna Kea observatories (S. Kulkarni, private
communication).

\begin{figure}
\plotone{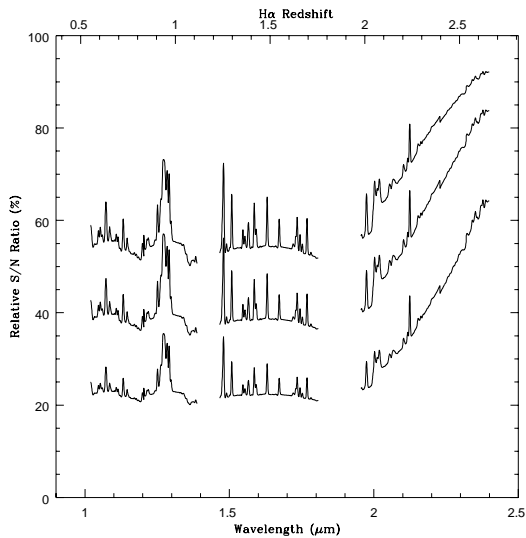}
\caption{Relative S/N ratios achieved binning data from 50, 35, and 20 mas lenslets (solid curves, top to bottom
respectively) to 100 mas, as compared
to data taken using 100 mas lenslets.
Redshifts for which H$\alpha$ emission falls at a particular wavelength are indicated along the top axis of the plot.}
\label{Binning.fig}
\end{figure}

Ideally, observations could be conducted using the finest available lenslet scale for which the field of 
view covers the entire target, since (even if the signal
is too low to usefully measure on such a fine scale) the results may be spatially binned to coarser scales.
In practice however detector noise often imposes severe penalties on fine sampling scales, introducing
noise into the binned sample.
Using our characterization of the Keck/OSIRIS
noise budget, we assess the utility of such binning as a function of wavelength by plotting the fractional change in the
S/N ratio obtained by binning data from the three smaller lenslet scales to 100 mas (Fig. \ref{Binning.fig}).  These calculations adopt
an individual exposure time of 900 seconds, and assume that the flux from the desired target is negligible compared 
to the background flux.
We note that for wavelengths longward of 2.3 $\micron$ the S/N ratios achieved for binning 50 mas lenslets to 100 mas scales is about 90\% of that
achived using actual 100 mas sampling, suggesting that for long wavelength observations it may be worthwhile to use small lenslets
even for faint
sources (although finer sampling will come at the cost of decreased field-of-view).

\subsection{Observing Star-Forming Regions at Intermediate to High Redshifts}

To assess the anticipated performance of OSIRIS we estimate the S/N ratios expected for observations of
star-forming regions at redshifts $z =$ 0.5 - 2.5,
considering the cases of a spatially unresolved (i.e. pointlike), isolated star-forming region, and a spatially extended source with a physical diameter of 1 kpc.
In each case it is assumed that the star-formation rate is 1 $M_{\odot}$ yr$^{-1}$ (assuming a Salpeter IMF this corresponds to 
an H$\alpha$ luminosity of $1.27 \times 10^{41}$ erg s$^{-1}$
from Kennicutt 1998), and that the velocity dispersion is negligible within a single spatial sample.
In each of these cases 
both simulated H$\alpha$ and [O III] $\lambda 5007$ emission lines (assuming [O III]/H$\alpha$ $\sim$ 1)
are artificially observed to calculate the expected S/N ratio expected in the peak
lenslet at sampling
scales of 100, 50, 35, and 20 mas.  These simulations take into account the redshift dependence of flux and angular diameter, 
and the wavelength dependence of the
atmospheric seeing\footnote{$\Theta_{\rm seeing} = \Theta_{\rm V} \times (\lambda/5500$ \AA$)^{-0.2}$, 
where visual-band seeing $\Theta_{\rm V} =$ 0.7 arcseconds.},  diffraction limit,
sky brightness, and AO correction strehl (assuming a fixed wavefront error $\sigma_{\rm wfe} = 337$ nm).

\begin{figure}
\plotone{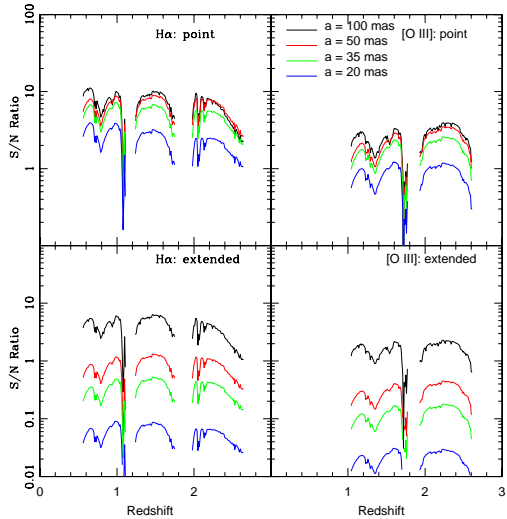}
\caption{Expected signal-to-noise ratios for 1 hour observations of point and extended (1 kpc diameter) sources with a star formation rate of 1 $M_{\odot}$ yr$^{-1}$ using different
OSIRIS lenslet scales ($a$).  The line flux ratio [O III] ($\lambda5007$)/H$\alpha$ is taken
to be unity, and the velocity dispersion in each spatial sample is assumed to be negligible.}
\label{snz.fig}
\end{figure}

Figure \ref{snz.fig} plots the results as a function of redshift for each of the available OSIRIS lenslet scales.  In general the shape of the curves
reflects the adopted atmospheric and spectrograph 
transmission spectra, with gaps where atmospheric transmission is particularly poor 
(OSIRIS filters are designed to cut out these wavelength
intervals).  The monotonic decrease in the S/N ratio for observations of H$\alpha$ at redshift $z > 2.1$ however is due to the rapidly rising thermal contribution
to the background radiation in the $K$-band.
We note that for point-like sources virtually the same signal level is 
obtained using both 100 mas and 50 mas sampling scales since nearly the same amount of integrated light
is contained within the lenslet, while losses begin to occur on
scales of 35 and 20 mas which are below the Keck diffraction limit.  In contrast, 
for extended sources the signal obtained using smaller lenslets is considerably reduced since
the lenslets are gathering light from a smaller effective area.\footnote{Recall that spillover from surrounding lenslets is assumed to be a source
of noise.  In localized areas, across which the physical properties of the source change very little, this spillover light from neighboring lenslets may instead boost
the observed signal (particularly for small lenslet scales), resulting in slightly higher S/N ratios than those conservative estimates given here.}

Observations of the rest-frame optical spectra of redshift $z \sim 2$ and $z \sim 3$ Lyman-Break Galaxies 
(LBGs, e.g. Pettini et al. 2001, Erb et al. 2003) have demonstrated that these galaxies
have strong nebular emission lines, particularly [O III] $\lambda$5007 for which the line flux ratio [O III]/H$\alpha$ can be close
to unity\footnote{[O III]/H$\alpha$ has not been directly measured, but the fluxes are often comparable.} 
for relatively metal-poor ($<$ 0.5 solar) sources.  Since the thermal background rises rapidly in the $K$ band 
(Fig. \ref{background.fig}), this strong [O III] emission can present
an appealing observational alternative to H$\alpha$ since it falls in the $H$ band (which has a lower thermal background) for redshifts $z \approx 1.9 - 2.6$.
However the exponential decrease of strehl with shorter wavelengths, paired with the fact that detector dark current rather than
thermal background radiation tends to dominate the error budget at most wavelengths and sampling scales, 
means that as shown in Figure \ref{snz.fig} (right panels) it 
will still be most efficient
to observe H$\alpha$ emission for all except those sources at $z > 2.5$, or at redshifts for which H$\alpha$ emission is severely attenuated by the atmosphere.

We conclude for redshift
$z \sim 2 - 2.6$ star-forming galaxies (which have typical uncorrected H$\alpha$ star-formation rates of $\sim$ 16 $M_{\odot}$ yr$^{-1}$, Erb et al. 2003) that
100 mas sampling of H$\alpha$ emission with OSIRIS should in general yield fairly comprehensive spatial coverage
with an integration time of a few hours (see also \S 3.3).  However,
background radiation and detector dark current will largely be prohibitive for obtaining spectroscopic data on scales of 50 mas
or below.  
As described in \S 3.1.2 though, it may nonetheless be desirable to use such fine sampling scales since: (1) Coarser
sampling scales can be reconstructed by binning data, often with relatively little loss of signal strength,
and (2) If emission regions visible in the ACS data resolve into point-like sources (e.g. giant H II regions
or super-star clusters) on scales below the HST-ACS resolution limit the signal on small scales may
be larger than anticipated.

\subsection{Signal-to-Noise Ratio Maps}
In order to reliably
reconstruct the velocity profile of a galaxy it will be necessary to obtain spectra with high S/N ratios over many lenslets.  Using the tools developed
so far we generate theoretical S/N ratio maps for four of the $z \sim 2 - 2.6$ starforming galaxies observed by Erb et al. (2004) to illustrate the spatial extent of the data
which Keck/OSIRIS will be capable of providing.

\begin{deluxetable}{cccc}
\tablecolumns{4}
\tablewidth{0pc}
\tabletypesize{\scriptsize}
\tablecaption{Parameters of Selected Galaxies\tablenotemark{a}}
\tablehead{
\colhead{Galaxy} & \colhead{$z_{\rm H\alpha}$\tablenotemark{b}} & \colhead{$F_{\rm H\alpha}$\tablenotemark{c}} & \colhead{$\sigma_{\rm v}$}\tablenotemark{d}}
\startdata
BX 1311 & 2.4843 & 16.0 & 88\nl
BX 1332 & 2.2136 & 8.8 & 54\nl
BX 1397 & 2.1332 & 10.6 & 123\nl
BX 1479 & 2.3745 & 5.0 & 46\nl
\enddata
\tablenotetext{a}{All data are from Erb et al. (2004).}
\tablenotetext{b}{Vacuum heliocentric redshift of H$\alpha$ emission line.}
\tablenotetext{c}{H$\alpha$ emission line flux in units of $10^{-17}$ erg s$^{-1}$ cm$^{-2}$.  Note these estimates have been corrected for slit aperture
and non-photometric conditions.}
\tablenotetext{d}{One-dimensional velocity dispersion of H$\alpha$ emission in units of km s$^{-1}$}
\label{Galaxies.table}
\end{deluxetable}

Using the instrument parameters outlined previously in Table \ref{Detector.table}, we simulate observations of BX galaxies 1311, 1332, 1397, and 1479
drawn from the
Steidel et al. (2004) catalog.  These 
four sources were selected because they all have long-slit H$\alpha$ spectra, two of them (BX 1332, 1397) 
with actual tilted velocity curves, and generally have
elongated or multiple-component morphologies.
H$\alpha$ fluxes for these galaxies are estimated using lower limits to the fluxes measured by Erb et al. (2004); motivated by narrowband imaging
and continuum spectroscopy of similar sources we estimate a factor of $\sim$ 2
correction to the Erb et al. (2004) values to account for slit losses and non-photometric observing conditions 
(see Table \ref{Galaxies.table} for a summary of these
corrected flux values).
The average H$\alpha$ flux of the BX galaxy sample is roughly $\sim 6 \times 10^{-17}$
erg s$^{-1}$ cm$^{-2}$ (Erb et al. 2005), and therefore while BX 1332 and 1397 are slightly brighter
than average we note that BX 1479 may be taken as roughly representative of the typical flux of the sample.
These H$\alpha$ fluxes are
used to scale the ACS $B$-band (F450W) images to generate H$\alpha$ flux maps as described in \S 2.1.2.
In each case we assume that the velocity dispersion is negligible within a single spatial sample.\footnote{This assumption will be valid so long as
the velocity dispersion
$\sigma_{\rm v}$ within the sample is less than $c/{\cal R}/2.4$ ($\approx 30$ km s$^{-1}$ for spectral resolution ${\cal R} = 3800$).  This translates to a total velocity
dispersion for a source of angular size 1 arcsecond (with a smoothly varying velocity field) of about 300 km s$^{-1}$ using 100 mas sampling, and 1500 km s$^{-1}$
using 20 mas sampling.}

\begin{figure}
\plotone{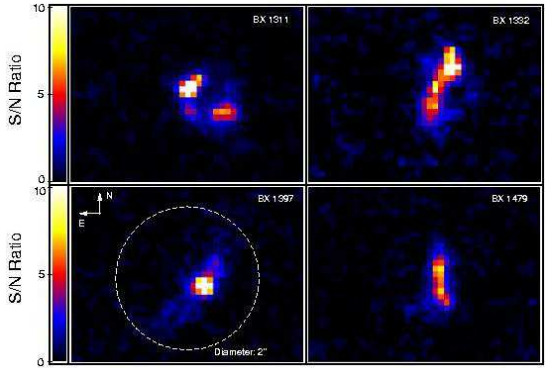}
\caption{Estimated S/N ratio maps for $z \sim 2 - 2.6$ star forming galaxies BX 1311, 1332, 1397, and 1479.  Simulated observations use 100 mas lenslets and image H$\alpha$
 emission for 4 hours of total integration time.  The velocity dispersion in each spatial sample is assumed to be negligible.}
\label{SNmap1.fig}
\end{figure}

\begin{figure}
\plotone{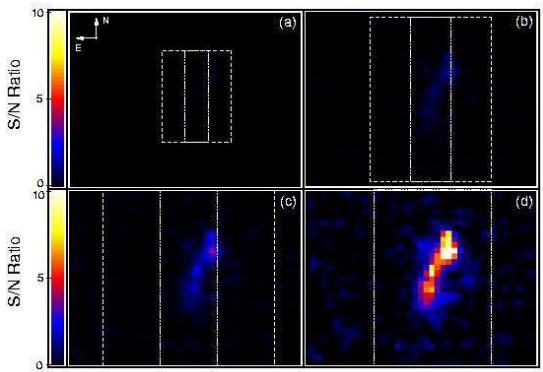}
\caption{As Figure \ref{SNmap1.fig}, but plots show S/N ratio maps for BX 1332 using lenslet scales of 20, 35, 50, and 100 mas (panels a.-d. respectively).
OSIRIS broadband and narrowband fields of view (Table \ref{FOV.table}) for each lenslet scale are overplotted (dotted and dashed lines respectively)
for comparison.}
\label{SNmap2.fig}
\end{figure}

S/N ratio maps of these four galaxies are given in Figure \ref{SNmap1.fig}, and show that in a four hour integration we expect to obtain
numerous spectra with S/N ratios $\gtrsim$ 5 (red pixels) across the source if a lenslet scale of 100 mas is used (as expected 
on the basis of Fig. \ref{snz.fig}).
In comparison, Figure \ref{SNmap2.fig} illustrates the S/N ratios expected in the same
integration time for BX 1332 using the four available lenslet scales of 20, 35, 50, and 100 mas (panels a - d respectively).
Clearly, 20 and 35 mas sampling scales are unlikely to be of use for observations of $z \sim 2$ galaxies since peak S/N ratios for 
emission line detection are less than or
of order unity in a four hour exposure, indicating that the integration times required
to achieve detectable signals 
are prohibitively large.  As demonstrated by panel (c), 50 mas sampling scales could nonetheless prove of use
when spatially binned, for brighter sources, or for those
sources in which emission regions resolve out into identifiable point sources.

In the event that the assumption of negligible velocity dispersion within a single lenslet were to break down (as when observing an AGN),
the line emission would be spread over a greater number of detector pixels, increasing the noise in the detection
and resulting in S/N ratios somewhat lower than found for unresolved line emission.

\subsection{Recovery of Velocity Structure}
As described in \S 2.2, detecting line emission in a given lenslet does not necessarily mean that it will be possible to measure
a value of the velocity centroid to within an accuracy necessary for studying velocity substructure (i.e. tens of km s$^{-1}$) since severe
contamination of the light in a given lenslet by spillover from a bright nearby component may mask the velocity signature.  We therefore
test how accurately we can expect to reconstruct the velocity field of BX 1332 given two possible kinematic models for the system.

\begin{figure}
\plotone{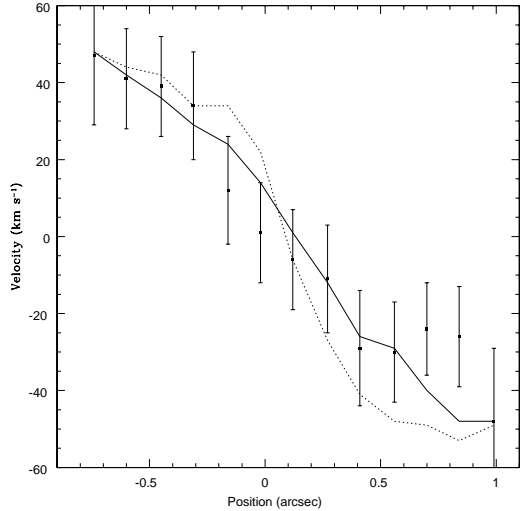}
\caption{Observed Keck/NIRSPEC velocity curve of BX 1332 (filled squares, from Erb et al. 2004) overplotted with simulated Keck/NIRSPEC observations 
assuming rotating disk kinematics (solid line) and a bimodal velocity distribution (dotted line).  Model velocities have been uniformly scaled along the vertical
axis to precisely match the dynamic range of the observational data.}
\label{VelCurves.fig}
\end{figure}

\begin{figure}
\plotone{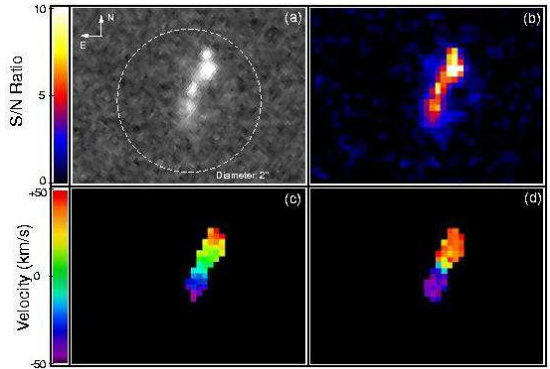}
\caption{(a) HST-ACS $B$-band image of BX 1332 from the GOODS-N survey.  Panel (b) reproduces the S/N ratio map for spectroscopic detection of the galaxy,
and panels (c) \& (d) respectively show the velocity profiles reconstructed from simulated observational data for rotating disk and bimodal
velocity structures.  Panels (c) \& (d) plot only data from those lenslets with greater than 3$\sigma$ detections of emission line flux, and for which the recovered
velocity is within 1000 km s$^{-1}$ of the systemic redshift.}
\label{VelFig.fig}
\end{figure}

In a previous study using long-slit Keck/NIRSPEC spectroscopy Erb et al. (2004) observed 
a tilted velocity curve along the major axis of BX 1332 (Fig. \ref{VelCurves.fig}, filled squares), but the high degree of correlation among
their data points due to atmospheric seeing ($\sim 0.8$ arcsec) meant it was uncertain whether the tilted velocity curve was due to ordered 
disk rotation,
or simply to two clumps merging or otherwise passing each other at some relative velocity.  We therefore construct two kinematic models for this 
source: 1) A disk exhibiting solid-body
rotation for which the size, inclination, and position angle\footnote{The model disk has a diameter of 1.4 arcseconds (as seen face-on, roughly corresponding
to a linear size of 12 kpc), a position
angle of -20$^{\circ}$, and an inclination to the line of sight $i = 70^{\circ}$.  The projected radial velocity is set to be 50 km s$^{-1}$ at the edge of the disk
(Erb et al. 2004), 
corresponding to a disk of mass $M = 4 \times 10^{9} M_{\odot}$.} are chosen so that the projected 
ellipsoid roughly traces the contours where the S/N ratio
is equal to 3 in Figure \ref{SNmap2.fig} (panel d.), 2) A bimodal velocity distribution in which the velocities of the northwest and southeast regions
are separated by the total observed velocity gradient ($\sim$ 100 km s$^{-1}$).
In each case, the intrinsic velocity dispersion within a given spatial sample is assumed to be negligible.
These velocity maps are superimposed on the line flux map (generated as described in \S 3.3) on a pixel-to-pixel basis.
While there are many uncertainties
and degeneracies in this method the exact set of parameters used for the two models is of little consequence
since we are primarily interested in how well we can distinguish between two fundamentally distinct kinematic models on the 
basis of velocity maps reconstructed
from simulated observational data (\S 2.1.4 and 2.2).
This is illustrated in Figure \ref{VelFig.fig}, which plots the HST-ACS $B$-band image of BX 1332 (panel a.), the S/N ratio map for spectral
line detection (panel b.), and the velocity map recovered from a 4-hour exposure using the techniques described in \S 2.2 for both rotating disk and bimodal
velocity models (panels c. \& d. respectively).

Based on long-slit Keck/NIRSPEC observations, it would not be possible to distinguish these two kinematic models.
Adopting a rough model of the NIRSPEC spectrograph with spectral resolution ${\cal R} = 1400$ in the $K$ band, Figure \ref{VelCurves.fig} overplots
the Erb et al. (2004) observational data with model velocity curves recovered from simulations assuming a rotating disk model (solid line) and
a bimodal velocity model (dotted line).  Although these two model velocity curves are generally quite similar, a chi-squared analysis prefers the disk model
over the bimodal model with a chi-square probability function of 0.02 compared to $10^{-10}$.  However, this distinction is deceptive, since a bimodal velocity
model with identical shape but 80\% the dynamic range in velocity produces a chi-square fit as good as for the disk model.  Since the
velocity amplitude of the source is uncertain to within about 20\%, we conclude that
both rotating disk and bimodal velocity models are equally consistent with the
NIRSPEC velocity curve.

In contrast, with IFS, adaptive optics, and increased spectral resolution, simulations of
OSIRIS were able to clearly distinguish between the two models: in panel (c) of Figure \ref{VelFig.fig} the velocity map shows a smooth
transition from redshift to blueshift (relative to systemic) over the entirety of the source, while in panel (d) only a few pixels on the boundary between
the two clumps give intermediate recovered velocities (since residual 
atmospheric seeing blurs the light from the two clumps into a slightly broader emission feature
at close to the systemic redshift).
We therefore conclude that OSIRIS will in some cases be capable
of distinguishing between rotating disks 
and two-component mergers previously indistinguishable with long-slit spectrographs alone.
However, BX 1332 is a particularly extended source, which makes it more amenable to study than most galaxies at similar redshifts 
and for faint, compact, or
particularly disordered and chaotic systems obtaining accurate kinematic information for component features may prove challenging
since residual atmospheric seeing may blur the combined emission lines into broader emission features.

A comprehensive study of velocity recovery from a wide variety of kinematic models using OSIRIS modeling is underway, and will hopefully
determine general classes of galaxies which OSIRIS is
capable of distinguishing between on the basis of observed 2D velocity information.
In future, the ability to draw such distinctions may allow us to test the
predictions of heirarchical formation theory by quantifying the relative star formation rates in massive
disks versus in proto-galactic mergers at high redshift.

\subsection{H$\alpha$ Morphologies and Photometry}
In addition to providing valuable kinematic information, OSIRIS stands to provide high quality broadband, narrowband, and emission-line morphologies and photometry
of galaxies both at high redshift and in the local universe.
Since roughly 80\% of the wavelength range covered by the instrument is free of atmospheric OH emission, it will
be possible to sum only those spectral channels free of contamination to reach effective backgrounds up to 3 magnitudes fainter 
(in the $H$-band) than possible
with standard photometry.  Narrowband photometry will thus in a sense come ``free'' with kinematic studies, and the observer will be free post-observing to use
the data cube to create any of a number of desirable 
data products.  Since for photometric purposes all emission in a given lenslet may be summed regardless of small
velocity shifts, photometry of $z \sim 2 - 2.6$ star-forming galaxies will be possible on scales of at least 50 mas, and possibly below.
Such photometric and morphological data will present complementary information to the kinematic data, and will likely be of use in interpreting kinematic results.

\section{PHYSICAL LIMITATIONS}

Having explored the scientific promise of OSIRIS for the $z \sim 2$ galaxy sample
we extend our discussion to consider basic design specifications for an arbitrary ground-based telescope
and their impact on possible targets for integral-field spectroscopy.  As the range of parameter space is too broad to cover fully, we restrict ourselves
to considering an OSIRIS-like IFS with mechanical and electrical specifications as detailed in \S 3.1.
This study will therefore be specific to OSIRIS-type spectrographs and Mauna Kea-like atmospheric conditions but will indicate general trends.

\subsection{Limiting Line Fluxes: Sampling Scales}

\begin{figure}
\plotone{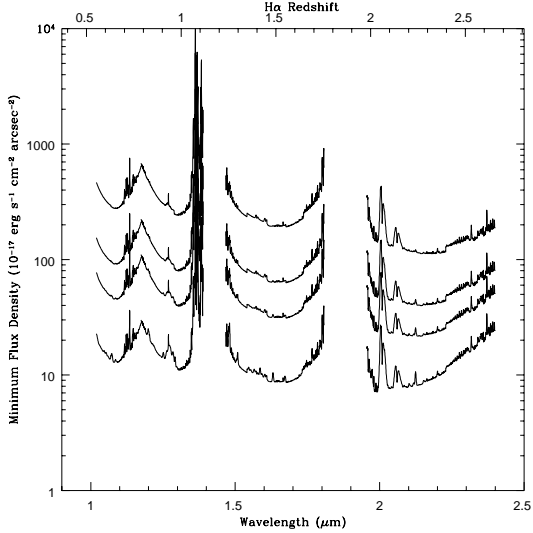}
\caption{Limiting line flux density (surface brightness) for a 5$\sigma$ detection in one hour of integration using Keck/OSIRIS.  
Curves are for (top to bottom) lenslet scales
of $a =$ 20, 35, 50, and 100 mas respectively.  Line emission is assumed to be spectrally unresolved.}
\label{limlenslet.fig}
\end{figure}

We begin by considering the limiting line fluxes which Keck/OSIRIS will be able to detect at the 5$\sigma$ level in one hour of integration.  Were detector noise
negligible, then for a ``large'' and uniformly bright extended source (i.e. a source for which strehl and PSF losses may be neglected) this line flux density would be 
roughly independent of lenslet sampling scale since
both sky background and line flux scale with the area $a^2$ of sky subtended by a lenslet.  However, as demonstrated previously in Figure \ref{mintime.fig},
detector noise dominates the noise budget at most wavelengths so while line flux still decreases with the lenslet area the noise
within a given lenslet remains approximately constant.  As indicated in Figure \ref{limlenslet.fig}, the limiting line flux therefore tends to increase
dramatically as lenslet size shrinks.  Unsurprisingly, this increase in limiting flux is less rapid for unresolved point sources.

\subsection{Limiting Line Fluxes: Telescope Aperture}

\begin{figure}
\plotone{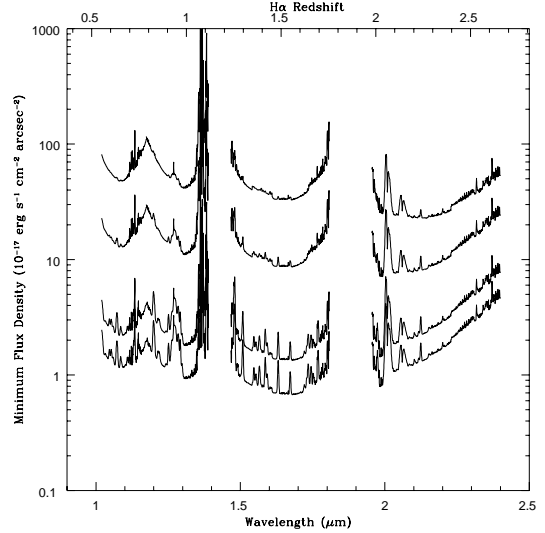}
\caption{As Figure \ref{limlenslet.fig}, but for $a =$ 100 mas lenslets using an
OSIRIS-like IFS on telescopes of assorted primary mirror
diameters ($D$).  Curves are for (top to bottom) telescopes with $D =$ 5, 10, 30, and 50 meters.}
\label{limtdiam.fig}
\end{figure}

Increasing telescope aperture for an AO-enabled telescope affects the limiting line flux both
by increasing the photon collection area and by concentrating the PSF of
the diffraction limited image,
effectively increasing signal in proportion to the primary mirror diameter $D^4$.
The net effect as a function of wavelength is shown in Figure
\ref{limtdiam.fig} for telescopes with $D = 5$, 10, 30, and 50 meters.
In addition to the improvements in limiting flux, 30-m and larger class telescopes will also permit
spatially resolved studies at finer angular scales than possible with current
generation instruments since the diffraction limit of a 30-m telescope is $\sim$ 18 mas in the $K$ band (roughly one third that of the Keck
telescope).

\begin{figure}
\plotone{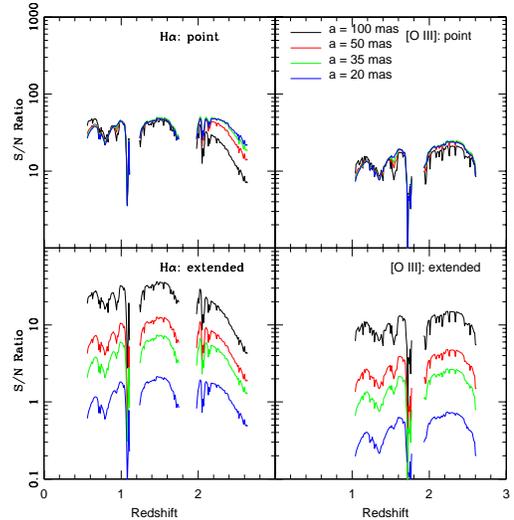}
\caption{As Figure \ref{snz.fig}, but for a thirty-meter class telescope with an OSIRIS-like IFS.}
\label{snzTMT.fig}
\end{figure}

To illustrate this point we repeat our earlier calculations of the expected S/N ratios for detection of emission from star forming regions
using a simulated thirty-meter class telescope (TMT)
in combination with an OSIRIS-like spectrograph.  As shown in Figure \ref{snzTMT.fig}, with this configuration
even spatially extended sources observed with lenslet sampling scales of 35 mas ($\sim$ 300 pc at redshift $z = 2.5$)
provide reasonable detections for spectroscopically unresolved emission
lines at redshifts up to $z = 2.5$ in only an hour of integration.
On scales of 100 mas, TMT is likely to provide H$\alpha$ spectra for point-like regions forming stars at a rate of 1 $M_{\odot}$ yr$^{-1}$
with a peak S/N ratio of 50+, permitting detailed
study of even the fainter components of target galaxies.

\subsection{Limiting Line Fluxes: Cryogenic Considerations}

\begin{figure}
\plotone{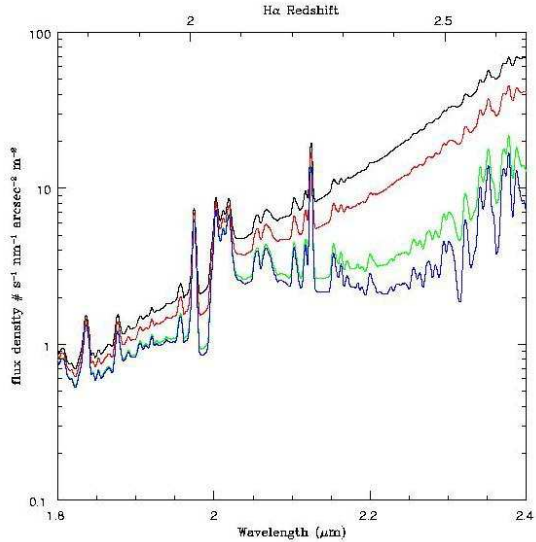}
\caption{Model effective $K$-band background spectra for: (1) 35\% emissivity in AO optics and 20\% emissivity in telescope optics at temperature $T = 270 K$ (black line,
as Figure \ref{background.fig}), (2) 35\% emissivity in AO at $T = 270 K$ and 0\% telescope emissivity (red line), (3) 35\% emissivity in AO cryogenically cooled
to $T = 250 K$ and 0\% telescope emissivity (green line), (4) 35\% emissivity in AO cryogenically cooled to $T = 77 K$ and 0\% telescope emissivity (blue
line).  All models incorporate the Gemini model for zodiacal light and atmospheric emission, for further commentary see Figure \ref{background.fig}.}
\label{kbackground.fig}
\end{figure}

As discussed in \S 3.2 thermal emission dominates the $K$-band background, particularly for large aperture telescopes.
This emission comes from telescope primary and secondary mirrors, re-imaging optics in
the AO system light path, and atmospheric contributions.  
Although the latter is uncontrollable for ground-based instruments, it is of obvious interest to attempt to minimize the contribution from optical components (Fig. \ref{kbackground.fig}).

\begin{figure}
\plotone{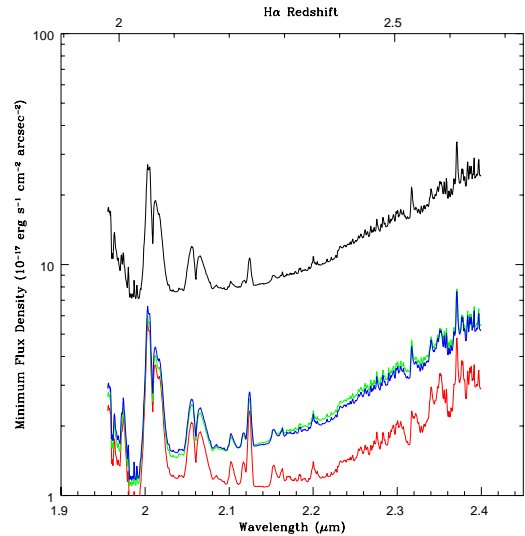}
\caption{Limiting line flux density in the $K$-band for a 5$\sigma$ detection in one hour of integration using an OSIRIS-like IFS on a 30-m class telescope (100 mas sampling).  
Curves are for: (1) 35\% emissivity in AO optics and 20\% emissivity in telescope optics at temperature $T = 270 K$
(black line),
(2) 35\% emissivity in AO at $T = 270 K$ and 0\% telescope emissivity (green line),
(3) 35\% emissivity in AO cryogenically cooled to $T = 250 K$ and 20\% telescope emissivity at temperature $T = 270 K$
(blue line),
(4) 35\% emissivity in AO cryogenically cooled to $T = 250 K$ and 0\% telescope emissivity (red line).
Line emission is assumed to be spectrally unresolved.}
\label{limcryo.fig}
\end{figure}

In Figure \ref{limcryo.fig} we plot the limiting line flux at $K$-band wavelengths for a 30-m class telescope operated under a range of thermal conditions.
Assuming it were possible to cryogenically cool
the AO bench to $T = -20^{\circ} C$, the thermal emission from the AO system would become largely negligible compared to that from the telescope and atmosphere,
pushing the limiting flux in the $K$-band a factor of three lower.  If, in addition, it were possible to achieve near-perfect 
reflectivity from the telescope mirrors, then the combined reduction in background flux would
reduce the limiting line flux by a factor of almost ten.

This improvement is most noticeable around wavelengths of 2.2 - 2.4 microns where the atmosphere is particularly transparent,
and we therefore gain the most from suppressing emission from optical components.  Longward of 2.4 microns however the improvement becomes less
pronounced as atmospheric emission begins to increase rapidly.

It is therefore an attractive goal for next-generation telescopes to cryogenically cool their adaptive optics system since a temperature change of only 20
degrees (coupled with increased mirror reflectivity)
may suffice to increase sensitivity in the $K$-band by almost an order of magnitude.  This 20 degrees of cooling would reduce 
thermal emission from the AO system to the level of the background
sky, so additional cooling (e.g. to liquid nitrogen temperatures $T = 77$ K) would not greatly decrease the background further.
Such a cooling system should not pose a great technological challenge (C. Max, private communication), although it would somewhat increase the cost
and complexity of such a telescope project.

\subsection{Spectral Resolution Requirements}
Moderate spectral resolution (${\cal R} \sim$ 3000) is necessary for spectroscopic studies of high redshift galaxies, largely so that individual atmospheric OH emission
lines may be suppressed, and in general the greater the resolution the greater the range of wavelengths for which it will be possible
to achieve inter-line background levels.  In addition, high spectral resolution aids in the successful recovery of velocity substructure,
and spectrograph resolution must be great enough that it is possible to 
measure velocities to within tens of km s$^{-1}$.  We note, however, that the benefits of increasing
spectral resolution do not continue indefinitely, but peak when emission lines in a lenslet are just barely unresolved, beyond which
detector noise begins to grow as the line emission is spread over a greater number of pixels.

\subsection{Benefits of Adaptive Optics}
Adaptive optics is a key component for integral-field observations of targets whose angular size is less than or comparable to the size of the seeing disk.
Most importantly, AO correction concentrates an appreciable fraction of emitted light into the PSF core of
the lenslet subtending the region from which that light was
emitted, greatly increasing the signal strength from point sources and permitting improved imaging of structures blurred together in
the seeing disk.  We note however that for slightly larger sources ($\gtrsim 4$ arcsec$^2$) atmospheric seeing may also be partially corrected for
using ground-layer adaptive optics (GLAO) techniques (e.g. Chun 2003)
or post-processing wavelet decomposition methods (e.g. Puech et al. 2004, Flores et al. 2004) instead of a conventional adaptive optics system.

One particularly useful observation for which AO correction will be essential
is the targeted observation of AGN with integral field spectrographs.  
In many sources the central AGN often outshines its host galaxy, which may be extremely difficult to
study if the AGN emission is blurred over its host in an atmospheric seeing halo.  With AO correction however, the AGN emission may be concentrated more
fully in its PSF core, decreasing the contamination in the seeing halo and possibly enabling an IFS to probe the emission from, and kinematics of, the host galaxy.

\section{SUMMARY}
We have outlined an observing strategy for infrared integral-field spectroscopy of high-redshift galaxies, and presented a formalism for simulating
observations with an arbitrary telescope plus lenslet spectrograph system.  These calculations may be generalized to predict the integration times
required to observe extended or compact sources both at high redshift and in the local universe.

We have determined 5$\sigma$ limiting line flux densities as a function of wavelength
for typical AO-equipped integral-field spectrographs
mounted on telescopes with primary mirror diameters ranging
from 5 to 50 meters.  Using 100 mas angular sampling,
these limiting flux densities are typically $\sim 10^{-16}$ erg s$^{-1}$ cm$^{-2}$ arcsec$^{-2}$ for 10-m class telescopes,
and $\sim 10^{-17}$ erg s$^{-1}$ cm$^{-2}$ arcsec$^{-2}$ for 50-m class telescopes. 
These limiting line fluxes increase by a factor of $\sim$ 3 for each factor of $\sim$ 2 that the angular lenslet scale is decreased.
Realistic models of the infrared background flux are taken into account in these calculations, and include blackbody emission from both
the atmosphere and telescope optics and atmospheric OH line emission.
Warm adaptive optics systems contribute significantly to the $K$-band background flux, and cryogenic cooling to $T = 250$ K
may reduce the effective background by as much as a factor of ten.

Using parameters characteristic of the Keck/OSIRIS spectrograph
we have predicted S/N ratios for detection of line emission from star-forming regions at redshifts $z = 0.5 - 2.5$ 
with angular lenslet scales of 20, 35, 50, and 100 mas, and find that OSIRIS can observe H$\alpha$ emission from regions forming stars at a rate of about
1 $M_{\odot}$ yr$^{-1}$ per 100 mas angular grid square with a signal-to-noise ratio S/N $\sim 10$.
Simulated integral-field spectroscopy of $z \sim 2 - 2.6$ star-forming galaxies in the GOODS-N field
indicates that OSIRIS can observe H$\alpha$ emission from these sources with an angular resolution of 100 mas ($\sim$ 1 kiloparsec at redshift $z \sim 2$)
and signal-to-noise ratios between 3 and 20,
and will be able to produce two-dimensional velocity maps for these galaxies relative to the systemic redshift.
These simulated velocity maps suggest that OSIRIS will be able to distinguish between some disparate kinematic models permitted by ambiguous long-slit
spectroscopic data alone, and may distinguish between mature disk kinematics and active mergers for spatially extended, 
bright galaxies at these redshifts.

Further observational constraints on the H$\alpha$ morphologies of the $z \sim 2 - 2.6$ galaxies from AO-corrected narrow-band photometry
will help determine whether sources may resolve into individual star forming regions on scales less than a kiloparsec, permitting more detailed
simulations and kinematic studies.
The morphological and velocity structure of these galaxies at such small scales may be
observable with OSIRIS for particularly bright sources,
and will certainly be accessible to future studies using next-generation 30-m and 50-m class telescopes.

\acknowledgements
The authors thank James Larkin for numerous helpful discussions regarding the OSIRIS spectrograph,
and for kindly providing the results of early commissioning tests of the instrument.  DRL also thanks 
D. LeMignant, K. Matthews, and K. Taylor for helpful discussions.  CCS, DKE, and DRL have been supported by grant
AST03-07263 from the US National Science Foundation, and AR grant HST-AR 10311 from the Space Telescope Science Institute.

\end{document}